\documentstyle[12pt]{article}
\advance\textheight by \topskip
\begin{document}
\begin{flushright}
SU-ITP-93-17\\
astro-ph/9307002\\
July 1993
\end{flushright}
\vspace{1 cm}
\thispagestyle{empty}
\begin{center}
{\Large\bf    HYBRID INFLATION}\\
\vskip 2 cm
{\bf Andrei Linde} \footnote{On leave from: Lebedev
Physical Institute, Moscow.\, \,
E-mail: linde@physics.stanford.edu}\\
\vskip .3cm
 Department of Physics, Stanford University, Stanford, CA 94305\\
\vskip 2 cm
{\large ABSTRACT}
\end{center}
\begin{quotation}

Usually inflation ends either by a  slow rolling of the inflaton
field, which gradually becomes faster and faster, or by a first-order
phase transition. We describe a model where inflation ends in a
different way,  due to a very rapid rolling  (`waterfall') of  a
scalar field $\sigma$ triggered by another scalar field $\phi$. This
model looks as a hybrid of  chaotic inflation with  $V(\phi) =
{m^2\phi^2\over 2}$ and the usual theory with spontaneous symmetry
breaking with  $V(\sigma) = {1\over 4\lambda}(M^2
-\lambda\sigma^2)^2$. The last stages of inflation in this model are
supported not by the  inflaton  potential $V(\phi)$ but by the
`non-inflationary' potential $V(\sigma)$.\\

Another hybrid model to be discussed here  uses some building blocks
from extended inflation (Brans-Dicke theory), from new inflation
(phase transition  due to a non-minimal coupling of the inflaton
field  to gravity) and from chaotic inflation (the possibility of
inflation beginning at large as well as at small $\sigma$).  In the
simplest version of this scenario  inflation ends up   by  slow
rolling, thus avoiding the big-bubble problem of  extended inflation.

\end{quotation}

\vspace{1cm}
\centerline{PACS number~~~98.80.Cq}

\vfill \newpage
\vfill\eject

1.  There exist three  independent ways of classifying
inflationary models. The first classification deals with the initial
conditions
for inflation. The old and new inflationary models were based on the
assumption that the Universe from the very beginning was in a state
of thermal
equilibrium at an extremely high temperature, and that the inflaton
field
$\phi$ was in a state corresponding to the minimum of its
temperature-dependent effective potential $V(\phi)$ \cite{b15,b16}.
The main
idea of the
chaotic inflation  scenario was to study {\it all} possible initial
conditions in
the Universe, including those which describe the Universe outside of
the state
of thermal equilibrium, and the scalar field outside of the minimum
of $V(\phi)$ \cite{b17}. This scenario includes the
possibility of  new
inflation from the state in a thermal   equilibrium, but it contains
 many
other possibilities as well. Therefore it can be realized in a much
greater variety of
models than the new inflationary Universe scenario. In fact, at
present the
idea of thermal beginning is almost completely abandoned, and   all
realistic models of inflation from the point of view of the first
classification
are of the chaotic inflation type \cite{MyBook}.

The second classification describes various regimes which are
possible
during inflation: quasiexponential inflation, power law inflation,
etc.  This
classification is absolutely independent of the issue of initial
conditions.
Therefore it does not make any sense to compare, say, power law
inflation and
chaotic inflation, and to oppose  them to each other. For example, in
\cite{ExpChaot} it was pointed out that the chaotic inflation
scenario, as
distinct from the new inflationary Universe scenario, can be realized
in the
theories with the effective potential $e^{\alpha\phi}$ for $\alpha
\ll \sqrt{16\pi}$. Meanwhile, in \cite{Exp} it was shown that this
inflation is
power law. Thus, the inflationary Universe scenario in the theory
$e^{\alpha\phi}$ describes  chaotic power law inflation.

Finally, the third
classification is related to the way inflation ends.  There are two
possibilities
extensively discussed in the literature: slow rollover versus the
first-order
phase transition. The models of the first class describe  slow
rolling of the
inflaton field $\phi$, which gradually becomes faster and faster. A
particular   model of this type is  chaotic inflation in the theories
$\phi^n$.
The models of the second class should contain at least two scalar
fields,
$\phi$ and $\sigma$. They describe a strongly first-order phase
transition
with bubble production which is triggered by the slow rolling of the
field
$\phi$. One of the popular models of this type is the extended
inflation
scenario \cite{b90}, which is a combination of  the  Brans-Dicke
theory and the old
inflationary scenario. There exist
other versions of the first-order scenario with two scalar fields,
which do not
require any modifications of the Einstein theory of gravity, see e.g.
\cite{EtExInf}.

In the beginning it was assumed that the bubbles formed during the
first-order
phase transition could be a useful ingredient of the theory of the
large scale structure
formation. However, later it was realized that one
should make considerable modifications of the original models in
order to
avoid disastrous consequences of the bubble production. According to
the most
recent modification \cite{CrittStein},   the bubble formation happens
only after
the end of inflation. In this case, the end of inflation occurs as in
the standard
slow-rollover scenario.  Therefore it would be interesting to find
out other
possible ways in which  inflation  may end in the models with several
different scalar fields. More generally, one may try to find out
other qualitatively new inflationary regimes which may appear  due to
a combined evolution of  several scalar fields.

Of course, one should not invent excessively complicated models
without
demonstrated need. However, sometimes  qualitatively different
inflationary regimes appear after  minor modifications of  the basic
inflationary models, or after making their hybrids.
For example,  in \cite{Axions} we proposed a very simple model of two
interacting scalar fields where  inflation may end  by a rapid
rolling of the
field $\sigma$ (`waterfall') triggered by the slow rolling of the
field $\phi$.
This regime differs both from the slow-rollover and the first-order
inflation.
By changing parameters of this model  one can
continuously interpolate between these two regimes. Therefore some
hybrid models of such type may share the best features of the
slow-rollover and the first-order models.

This model was only briefly introduced in \cite{Axions}. One of the
purposes of  the present paper is to discuss
this model  in a more detailed way.
Another  model to be discussed in this paper looks like a  hybrid of
the Brans-Dicke theory and new inflation. This model is similar to
extended inflation, but is does not suffer from the big-bubble
problem, since the phase transition which occurs in this model is
second order. Finally, we will discuss a model which looks like a
hybrid of the Brans-Dicke theory and chaotic inflation. In this model
the Universe after inflation becomes divided into exponentially large
domains with different values of the Planck mass $M_p$ and of the
amplitude of density perturbations ${\delta\rho\over\rho}$.

\

2.  We begin with the discussion of the hybrid inflation model
suggested in
\cite{Axions} in the context of the Einstein theory of gravity. The
effective
potential of this model is given by
\begin{equation}\label{hybrid}
V(\sigma,\phi) =  {1\over 4\lambda}(M^2-\lambda\sigma^2)^2
+ {m^2\over
2}\phi^2 + {g^2\over 2}\phi^2\sigma^2\ .
\end{equation}
Theories of this type were considered in \cite{KL}--\cite{HBKP}. The
main
difference between the models of refs. \cite{KL}--\cite{HBKP} and our
model is a specific choice of parameters, which allows the existence
of the
waterfall regime mentioned above. There is also another important
difference: we will
assume that the field $\sigma$ in this model is the Higgs field,
which  remains as
a physical degree of freedom after the
Higgs effect in an underlying gauge theory with spontaneous symmetry
breaking. This  field acquires only positive values, which
removes the possibility of domain wall formation in this theory.
Usually
it is rather dangerous to take the inflaton field interacting with
gauge fields, since its effective coupling constant $\lambda$ may
acquire large radiative corrections $\sim e^4$, where $e$ is the
gauge coupling constant. In our case this  problem does not appear
since density perturbations in our model remain small at rather large
$\lambda$, see below.

The effective
mass squared of the field $\sigma$ is equal to  $-M^2 + g^2\phi^2$.
Therefore
for $\phi > \phi_c = M/g$ the only   minimum of the effective
potential
$V(\sigma,\phi)$ is at $\sigma = 0$. The curvature of the effective
potential in the $\sigma$-direction is much greater than in the
$\phi$-direction. Thus we expect that at the first stages of
expansion of the Universe the field $\sigma$ rolled down to $\sigma =
0$, whereas the field $\phi$ could remain large for a much longer
time.  For this reason we will consider the stage of inflation at
large $\phi$, with $\sigma = 0$.

At the moment when the inflaton field
$\phi$ becomes smaller than  $\phi_c = M/g$, the phase transition
with the
symmetry breaking occurs. If $m^2 \phi_c^2 = m^2M^2/g^2 \ll
M^4/\lambda$, the Hubble constant at the time of the phase transition
is given by
\begin{equation}\label{2}
H^2 = {2\pi M^4 \over 3 \lambda M^2_p} \ .
\end{equation}
Thus we will assume that $M^2 \gg {\lambda m^2\over g^2}$.   We will
assume also that
$m^2 \ll H^2$, which gives
\begin{equation}\label{2a}
M^2\gg mM_p\sqrt{3\lambda  \over 2\pi} \ .
\end{equation}

One can easily verify, that, under this condition, the Universe at
$\phi > \phi_c$ undergoes a stage of inflation. In fact, inflation in
this model occurs even if $m^2$ is somewhat greater than $H^2$. Note
that
inflation at its last stages is driven not by the energy density of
the inflaton
field $\phi$ but by the vacuum energy density $V(0,0) = {M^4\over
4\lambda}$, as in the new inflationary Universe scenario. This was
the reason why we called this model `hybrid inflation' in
\cite{Axions}.

Let us study the behavior of the fields $\phi$ and $\sigma$ after the
time
$\Delta t = H^{-1} = \sqrt{3\lambda\over 2\pi}\,{M_p\over M^2}$ from
the
moment $t_c$ when the field $\phi$ becomes equal to $\phi_c$. The
equation of motion of the field $\phi$ during inflation is
$3H\dot\phi = m^2\phi$.
Therefore during the time interval $\Delta t = H^{-1}$ the field
$\phi$
decreases from $\phi_c$ by $\Delta\phi = {m^2\phi_c\over 3 H^2}=
{\lambda m^2 M^2_p\over 2\pi g M^3}$. The absolute value of the
negative effective mass squared $-M^2 + g^2\phi^2$ of the field
$\sigma$ at that time becomes equal to
\begin{equation}\label{3}
M^2(\phi) =  {\lambda m^2 M^2_p\over \pi  M^2} \ .
\end{equation}
The  value of $M^2(\phi)$ is much greater than $H^2$ for
$M^3 \ll    \lambda   \ m  M^2_p$.
In this case the field $\sigma$ within the time $\Delta t \sim
H^{-1}$
rolls down to its minimum  at $\sigma(\phi) = M(\phi)/\sqrt\lambda$,
rapidly oscillates near it and loses its energy due to the expansion
of the
Universe.  However, the field cannot simply relax near this minimum,
since
the effective potential $V(\phi,\sigma)$ at $\sigma(\phi)$ has a
nonvanishing partial derivative
\begin{equation}\label{5}
 {\partial V\over \partial\phi} = {m^2\phi} + {g^2\phi M^2(\phi)\over
\lambda}  \ .
\end{equation}
One can easily check that the motion in this direction becomes very
fast and the field $\phi$ rolls to the minimum of its effective
potential within the time much smaller than $H^{-1}$  if
$M^3 \ll  {\sqrt \lambda\, g } \ m  M^2_p $.
Thus, under the specified conditions inflation ends in this theory
almost
instantaneously, as soon as the field $\phi$ reaches its critical
value $\phi_c = M/g$.

The amplitude of  adiabatic density perturbations produced in this
theory
can be estimated by standard methods \cite{MyBook} and is given by
\begin{equation}\label{E24}
\frac{\delta\rho}{\rho} = \frac{16\sqrt {6 \pi}}{5}\
     \ \frac{V^{3/2}}{M_p^3\, {\partial V \over \partial\phi}} ~ = ~
 \frac{16\sqrt {6 \pi}\left({M^4\over 4\lambda} + {m^2\phi^2\over
2}\right)^{3/2}}{5 M_p^3\, {m^2\phi}}\ .
\end{equation}
In the case $m^2 \ll H^2$ the scalar field $\phi$ does not change
substantially during the last 60 $e$-foldings (i.e. during the
interval $\Delta t \sim 60 H^{-1}$). In this case the amplitude of
density perturbations practically does not depend on scale, and is
given by
\begin{equation}\label{hybrid2}
{\delta\rho\over \rho} \sim  {2\sqrt{6\pi} g M^5\over 5\lambda
\sqrt{\lambda} M^3_p m^2}\ .
\end{equation}
The definition of  ${\delta\rho\over \rho}$ used in   \cite{MyBook}
corresponds to COBE data for  ${\delta\rho\over \rho} \sim 5\cdot
10^{-5}$. Dividing it by (\ref{2a}) with an account taken of
(\ref{hybrid2}) gives $M^3 \ll 5\cdot 10^{-5} \lambda g^{-1} m
M_p^2$. This means that the `waterfall conditions'
$M^3 \ll    \lambda   \ m  M^2_p$ and $M^3 \ll  {\sqrt \lambda\, g  }
\ m  M^2_p $ automatically follow from the conditions $m^2\ll H^2$
and ${\delta\rho\over \rho} \sim 5\cdot
10^{-5}$, unless the coupling constants $\lambda$ and $g$ are
extremely small. Therefore the waterfall regime is realized in this
model for a wide variety of  values of parameters $m, M, \lambda$ and
$g$ which lead to density perturbations $\sim 5\cdot 10^{-5}$.

To give a particular example, let us take $g^2 \sim \lambda
\sim10^{-1}$,
$m \sim 10^2$ GeV (electroweak scale). In this case all conditions
mentioned above are satisfied and ${\delta\rho\over \rho} \sim 5\cdot
10^{-5}$ for $M \sim  1.3 \cdot10^{11}$ GeV. In particular, we have
verified, by solving equations of motion for the fields $\phi$ and
$\sigma$ numerically, that inflation in this model ends up within the
time $\Delta t \ll H^{-1}$ after the field $\phi$ reaches its
critical value $\phi_c = M/g$. The value of the Hubble parameter
at the end of inflation is given by $H \sim 7\cdot 10^3$ GeV.  The
smallness of the Hubble constant at the end of inflation
makes it possible, in particular, to have a consistent scenario for
axions in
inflationary cosmology even if the axion mass is much smaller than
$10^{-5}$
eV \cite{Axions}. This model has some other distinctive
features. For example, the spectrum of perturbations generated in
this model may look as a power-law spectrum rapidly decreasing at
large wavelength $l$ \cite{LythLiddle}.

Indeed, at the last stages of inflation (for ${M^4\over 4\lambda} \gg
{m^2\phi^2\over 2}$) the field $\phi$ behaves as
\begin{equation}\label{xx}
\phi = \phi_c \cdot \exp{\Bigl(-{m^2(t-t_c)\over 3 H}\Bigr)}\ ,
\end{equation}
whereas the scale factor of the Universe grows exponentially, $a \sim
e^{Ht}$.  This leads to the following relation between  the
wavelength of perturbations $l$ and the value of the scalar field
$\phi$ at the moment when these perturbations were generated: $\phi
\sim \phi_c \left({l\over l_c}\right)^{m^2/3H^2}$. In this case
\begin{equation}\label{xxx}
\frac{\delta\rho}{\rho}  =    {2\sqrt{6\pi} g M^5\over 5\lambda
\sqrt{\lambda} M^3_p m^2} \cdot \left({l\over l_c}\right)^{-{m^2\over
3H^2}}\ ,
\end{equation}
which corresponds to the spectrum index $n = 1 + {2m^2\over 3H^2} = 1
+{\lambda m^2M_p^2\over \pi M^4}$. Note that this spectrum index is
greater than $1$, which is a very unusual feature. For the values of
$m, M, \lambda$ and $g$ considered above, the deviation of $n$ from
$1$ is vanishingly small (which is also  very unusual). However, let
us take, for example, $\lambda = g = 1$, $M = 10^{15}$ GeV (grand
unification scale), and $m =   5\cdot 10^{10}$ GeV. In this case the
amplitude of perturbations at the end of inflation ($\phi = \phi_c$)
is equal to $4\cdot 10^{-4}$,  $n \sim 1.1$, and the amplitude of the
density perturbations drops to the desirable level  ${\delta\rho\over
\rho} \sim 5\cdot 10^{-5}$ on the galaxy scale ($l_g \sim l_c\cdot
e^{50}$). One may easily obtain models with even much larger $n$,
but this may be undesirable, since it may lead to  formation of many
small primordial black holes \cite{Polnarev}.

Note, that the decrease of  ${\delta\rho\over \rho}$ at large $l$ is
not unlimited.  At ${m^2\phi^2\over 2} > {M^4\over 4\lambda}$ the
spectrum begins growing again. Thus, the spectrum has a minimum on a
certain scale, corresponding to the minimum of expression
(\ref{E24}). This complicated shape of the spectrum appears in a very
natural way, without any need to design artificially bent potentials.

As we have seen, coupling constants in our model can be reasonably
large, and the range of possible values of masses $m$ and $M$ is
extremely wide. Thus, our model is very versatile. One should  make
sure, however, that the small effective mass of the
scalar field $\phi$ does not acquire large radiative corrections near
$\phi = \phi_c$. Hopefully this can be done in supersymmetric
theories with flat directions of the effective potential.

One can suggest many interesting generalizations of our model. For
example,  instead of the term ${m^2\phi^2\over 2}$ in (\ref{hybrid})
one can use the term ${\lambda_\phi\,\phi^4\over 4}$. In this case
one may have two disconnected stages of inflation. The  first stage
occurs at large $\phi$, as in the simplest version of chaotic
inflation scenario. This stage  ends at $\phi < M_p/3$, if $M^2 \ll
\lambda_\phi M_p^2$. Then the field  rapidly rolls down and
oscillates until the amplitude of its oscillations becomes smaller
than   $\phi \sim {M^2\over \lambda_\phi\, M_p}$. At this moment the
frequency  of oscillations $\sim \sqrt \lambda_\phi \phi$ becomes
smaller that the Hubble constant, and the second stage of inflation
begins. This stage of  inflation ends with the waterfall at $\phi_c =
M/g$. As was shown in \cite{Mukh}, in the models with two stages of
inflation with a break between them  the spectrum of density
perturbations may have a very reach and non-trivial structure.

\

3. Our second hybrid inflation model is very
similar to extended inflation, but it does not
lead to the first-order phase transition with bubble formation, which
is a definitive feature of the extended inflation scenario. The
corresponding action is
\begin{equation}\label{8}
S = \int d^4x \sqrt {-g} \  \Bigl[{{\phi^2 R}\over {8 \omega}}
-{\xi\over
2}\sigma^2 R +
{1\over 2} \partial_\mu  \phi \partial^\mu \phi +
{1\over 2} \partial_\mu  \sigma \partial^\mu \sigma - V(\sigma)
\Bigr]~.
\end{equation}
Here $\phi^2 = {{\omega\over 2 \pi }\Phi}$  is the
Brans-Dicke field; $\sigma$ is the   field which may exhibit
spontaneous symmetry breaking.  In the  extended inflation scenario
the second term in the action (\ref{8}) was absent. However, we
believe
that if one of the two scalar fields is non-minimally
coupled to gravity, one should allow for a similar coupling for
another field as
well.\footnote{After we proposed this model, we received a preprint
by Laycock
and Liddle \cite{Liddle}, where  a similar model was invented. We are
extremely
grateful to these authors for the discussion of their results prior
to
publication. Our understanding of the large $\xi$ limit of this model
strongly benefited from these discussions.}

We will consider effective potentials of the standard type,
$V(\sigma) = {1\over 4\lambda}(m^2-\lambda\sigma^2)^2$. From a
comparison with the standard Einstein theory it
follows that
the effective  Planck mass  in the theory (\ref{8}) depends on $\phi$
 and is given by
\begin{equation}\label{8a}
M^2_p(\phi) = 2 \pi\left({\phi^2\over \omega} - {4\sigma^2
\xi}\right) \ .
\end{equation}
We will try to find inflationary solutions in the theories with $w
\gg 1$,
and we will assume that initially $\sigma = 0$. The reasons why we
take $\sigma = 0$ at the beginning of inflation will be discussed in
the next sections. During inflation one can neglect
${\dot
\phi}\over \phi$ as compared with $H = {{\dot a} \over a}$ (where
$a(t)$ is
the scale factor of the Universe) and ${\ddot \phi}$ as compared with
$3H\dot
\phi$ and ${\dot\phi}^2$. In such a case equations for $a$ and $\phi$
in the theory (\ref{8}) take a very simple form:
\begin{equation}\label{x}
H = {{\dot a} \over a} = {2\over \phi} \sqrt{\omega  V(0)\over
3} = {m^2\over \phi}\sqrt{\omega\over  3\lambda} \  ,
\end{equation}
\begin{equation}\label{y}
3H\dot\phi = {4  V(0)\over \phi} = {m^4 \over \lambda\phi}\ .
\end{equation}
{}From these equations it follows that at large $t$
\begin{equation}\label{u}
a(t) = a_o t^\omega ~, ~~~ H(t) = {\omega\over t} \ ,~~~ \phi(t) =
{m^2 \, t\over\sqrt{3\omega\lambda}}~.\end{equation}
The effective mass squared of the field $\sigma$ is given by
\begin{equation}\label{1x}
m^2_\sigma = -m^2 + \xi\, R = -m^2 + 12\xi\,H^2 = -m^2 + {12\xi\,
\omega^2\over t^2} \ .
\end{equation}
This means that at small $t$ the effective potential including the
term
$-{\xi\over 2} \sigma^2R$ has a minimum at $\sigma = 0$.  Here one
may
consider  two limiting possibilities depending on the value of the
parameter $\xi$.

\

4.  Let us consider first the model with large $\xi$. In this case
our model  looks like a
hybrid of the Brans-Dicke theory and new inflation. In the very early
Universe the term $12 \xi H^2$ determines the effective mass squared
of the
scalar field $\sigma$. This mass squared is much greater than $H^2$,
which means that the field $\sigma$ rapidly rolls down to the state
$\sigma = 0$ (symmetry restoration), after which inflation begins.
Analogy between this scenario
and the new inflationary
Universe scenario becomes even more striking if one remembers that
the
term $12 \xi H^2$ can be formally represented as $48\pi^2 \xi\,
T_H^2$, where
$T_H = H/2\pi$ is the Hawking temperature in inflationary Universe.

One of the main problems of the new inflationary scenario is that
inflation in
this scenario may occur only at a density much smaller
than the Planck density. In this case a typical closed Universe of
the Planck size
$M_p^{-1}$ and with the Planck density $M^4_p$ collapses before
inflation
has any chance to occur. The probability of creation of a closed
Universe which can survive until the beginning of inflation is
exponentially small \cite{MyBook}. In our scenario this problem
is absent if the Universe
is created in a state with the fields $\phi$ and $\sigma$ related to
each other
by the condition $M^4_p(\phi,\sigma)\sim V(\sigma)$; there is no
exponential suppression of the probability of creation of the
Universe with such properties \cite{EtExInf}.

Thus, in this scenario inflation begins at $\sigma = 0$. Later on,
when the time $t$ becomes greater than $t_c$,
where
 \begin{equation}\label{2x}
t_c = w H^{-1}_c = {2\omega\sqrt{3\xi}\over m } \ , ~~~H_c = {m\over
2\sqrt{3\xi}}\ , ~~~ \phi_c =  2m\sqrt{\omega\xi \over
\lambda }~, \end{equation}
the effective mass squared of the field $\sigma$ becomes
negative, and spontaneous symmetry breaking occurs.

In the large $\xi$ limit, the absolute value of the (negative)
effective
mass squared of the field $\sigma$ becomes greater than $H_c^2$
within the
Hubble time $\Delta t = H_c^{-1}$ after $t_c$. At the moment $ t =
t_c
+H_c^{-1}$ the effective mass squared of the field $\sigma$ becomes
equal to
$m^2_\sigma = 12 \xi \omega^2(t^{-2} - t_c^{-2})
\approx - {24 \xi H^2_c/  \omega} $.
This quantity is much
greater than $H^2_c$ for $\xi > {\omega/24}$.
Under this condition, the field $\sigma$ begins growing with a
very large  speed, which suggests that inflation ends up almost
instantaneously. However, a more detailed investigation of this
question  shows
that the growth of the field $\sigma$ later slows down. The reason
for this
effect is rather nontrivial. When the field $\sigma$ grows, the
effective Planck
mass decreases, see eq. (\ref{8a}). This leads to an increase of the
Hubble
constant $H$, which slows down the rolling of the field $\sigma$.
This leads to
existence of an additional stage of inflation, which occurs
after the phase transition at $t = t_c$. In many cases this stage
proves to be relatively short \cite{Liddle}. Thus, in this model one
can
also have a `waterfall' regime, but this waterfall is much slower
than in the model (\ref{hybrid}).

This regime   has a peculiar feature, which deserves further
investigation. It is
 well known that the value of the Brans-Dicke field almost does not
change after
inflation \cite{Weinberg}. The change of this field during the last
stage of
inflation under the
condition $\xi > {\omega/24}$ is also very small. This means that
the present value of the
field $\phi$ in the first approximation is equal to $\phi_c$. On the
other hand,
the present value of the field $\sigma$ is given by $\sigma_0
= m/\sqrt\lambda$. According to (\ref{8a}), (\ref{2x}),   the
contributions to
$M_p$ from the fields $\phi$ and $\sigma$ in this approximation
cancel
each other,  $M_p = 0$! Thus, in order to obtain a finite answer for
the
gravitational constant $G = M_p^{-2}$ one should calculate the small
difference
$\Delta\phi$ between $\phi_c$ and the present value of the field
$\phi$:
\begin{equation}\label{Pl}
M_p =  4 m\,\sqrt{\pi\xi\over \lambda }\cdot \sqrt{\Delta\phi\over
\phi_c}
\sim 4 \sigma_0\, \sqrt{\pi\xi} \cdot \sqrt{\Delta\phi\over \phi_c} ~
{}.
\end{equation}
Alternatively, one may use
a different effective potential $V(\sigma)$, for which this peculiar
cancellation does not occur.
A more detailed investigation of the regime discussed above will be
contained
in the forthcoming paper by Laycock and Liddle \cite{Liddle}.

If the stage of inflation after the phase transition is very short,
one should take special care of production of topological defects in
this model.   Typically it is
very difficult to create heavy strings and monopoles after inflation.
However, it
is possible to produce them during inflationary phase transitions
\cite{KL,Olive}. If, for example, the   field $\sigma$ leads to
spontaneous
breakdown of an Abelian   symmetry, after the phase transition  in
our
model many heavy cosmic strings will be created. In fact,  cosmic
strings in
this model may be even too heavy. Indeed, according to (\ref{Pl}),
the scale of
symmetry breaking is $\sigma_0 = {M_p\over 4}\sqrt{\phi_c\over
\pi\xi\Delta\phi}$. This  is much greater than the desirable value
$\sigma_0 \sim 10^{-3} M_p$ \cite{Vilenkin}, unless
the parameter $\xi$ is extremely large. This means that if    string
production  is possible in our scenario, they may give excessively
large
contribution to the post-inflationary density perturbations.  There
exist
several different ways to reduce these perturbations to an acceptable
level.
One possibility is to consider the  field $\sigma$ of the type used
in the
standard theory of electroweak interactions, where neither stable
strings nor
stable domain walls or monopoles can be produced. One should keep in
mind, however, that  it may be necessary for the field $\sigma$ to be
a  gauge singlet, since in this model, unlike  in the first hybrid
inflation model (\ref{hybrid}), the coupling constant $\lambda$
should be very small unless one takes $\xi$ extremely large
\cite{Liddle}. The most radical way to get rid of topological defects
is to consider models where the last stage of inflation is long
enough. This can be achieved, e.g., in the model to be discussed
below.

\

5.  Let us consider the model (\ref{8}) with $V(\sigma) =
{\lambda\over 4}(\sigma^2 - {m^2\over \lambda})^2$ in the limit of
small
$\xi$. In this limit the phase transition becomes irrelevant since
the
correction $\xi R \sim 12\xi H^2$ to the effective mass of the field
$\sigma$
always remains much smaller than $H^2$. Therefore these corrections
cannot influence behavior of the field $\sigma$ during inflation in a
noticeable way.

For this reason we will take a step back and totally disregard the
term $-{\xi\over 2} R \sigma^2$ in the action (\ref{8}). At the first
glance, we are returning to
the standard extended inflation scenario. The difference, however, is
in the
choice of the effective potential $V(\sigma)$. In the extended
inflation
scenario the effective potential $V(\sigma)$ should have a local
minimum at
$\sigma = 0$. The simple potential we consider, which is a standard
potential
used in gauge theories with spontaneous symmetry breaking, does not
have this
extra minimum. In order to obtain such a minimum one should add some
cubic or
logarithmic corrections to $V(\sigma)$. Then one should tune these
corrections
to make the tunneling  suppressed, but not too strongly, since this
would make
the Planck constant exponentially large and the bubbles exponentially
big.
And, after all, one should either introduce a potential for the
Brans-Dicke
field or considerably modify the interaction of this field with
gravity \cite{CrittStein}. No such modification  is
required in our scenario.

At the first stages of inflation in our scenario the field $\sigma$
slowly rolls
down from some initial value $\sigma_{in} \ll \sigma_0$.  Until this
field grows
up approximately to $\sigma_0/2$, the effective potential $V(\sigma)$
remains almost unchanged, and its derivative is  given approximately
by
$-m^2\sigma$. Therefore at this stage eqs. (\ref{x})--(\ref{1x})
remain valid, and equation for the field $\sigma$ reads
$3 H \dot\sigma = {3\omega\over t}\, \dot\sigma = m^2\sigma$,
which gives
\begin{equation}\label{u2}
\sigma = \sigma_{in}\cdot \exp\left({m^2t^2\over 6\,\omega}\right)\ .
\end{equation}
This stage ends up   at the time $t \approx t_1$, when $\sigma(t_1) =
\sigma_0/2$. This gives
$t_1 = {\sqrt{6\omega}\over m}\,\log^{1/2}{\sigma_0\over
2\sigma_{in}}$, ~$ \phi_1 =
\sqrt 2\, \sigma_0\, \log^{1/2}{\sigma_0\over 2\sigma_{in}}$.
Exact values of $t_1$ and $\phi_1$ depend on $\sigma_{in}$, which may
take different values in dif\-ferent parts of the Universe. But the
situation
actually is even more complicated. As it was shown in \cite{Vil},
inflationary
Universe in the theories with the potential $V(\sigma) =  {1\over
4\lambda}(m^2 -\lambda\sigma^2)^2$ enters  regime
of self-reproduction at $\sigma \leq H$. This regime exists for $m^2
< H^2$. In
the context of our model this regime leads to formation of domains
with all
possible values of $\phi$ compatible with inflation at $\sigma = 0$.
The upper
boundary for inflation at small $\sigma$ (the condition $m^2 < H^2$)
is given
by $\phi < m\sqrt{w\over \lambda} = \sigma_0 \sqrt{\omega} $. In such
domains the  stage of classical growth of the field is very short,
and the
field $\phi$ remains almost unchanged when the field $\sigma$  grows
up to
$\sigma_1 \sim \sigma_0/2 = m/2\sqrt\lambda$. For a complete
investigation
of this question one should use stochastic approach to inflation. We
will return
to this problem in   \cite{BLL}. At the present moment we will just
keep in
mind that at the end of the first stage of inflation
the Brans-Dicke field may acquire different values in different parts
of the
Universe, in the range of  $\sigma_0 < \phi_1 < \sigma_0
\sqrt{\omega}
$. We will describe this effect by introducing a phenomenological
parameter $C$, such that $\phi_1 = C\sigma_0$\,,\, $1 < C <
\sqrt{\omega}$.

This means that at the end of the first stage, when the field
$\sigma$ grows up
to $\sigma \sim \sigma_0/2 = m/2\sqrt\lambda$, the square of the
Hubble
constant remains greater than the (positive) effective mass squared
$m^2(\sigma_0) = 2m^2$ of the field $\sigma$  near the minimum of its
effective potential at $\sigma = \sigma_0$. Therefore at that time
inflation
still continues. The effective potential $V(\sigma)$ near its minimum
can be
represented as  $V(\chi) = m^2 \chi^2$, where we made an obvious
change of variables, $\chi = \sigma_0 - \sigma$.

Fortunately, we already studied inflation in the Brans-Dicke theory
with this
potential, and all analytical solutions are known \cite{ExtChaot}:
\begin{equation}\label{n8}
\phi = A \cdot \sin \Bigl(B +  {m \over \sqrt {3\omega}} t\Bigr) \
,~~~~~
\chi \equiv \sigma_0 - \sigma  = {A\over \sqrt {2}}\cdot \cos \Bigl(B
+ {m \over
\sqrt {3\omega}} t\Bigr) \ ,
\end{equation}
  \begin{equation}\label{n10}
 a(t) =a(t_1)\cdot  {\left({\phi(t) \over
\phi_1}\right)}^\omega  = a_o
\cdot{\left({\sin (B + {m \over \sqrt {3\omega}} t)\over \sin
B}\right)^\omega} \ . \end{equation}
 Here $A$ and $B$ are some constants determined by initial
conditions. For
${m \over \sqrt {3\omega}} t > B$, $\omega \gg 1$, the last equation
describes the power law
inflation, $a(t) \sim t^\omega$.

Initial conditions for this
stage of inflation are determined by $\chi_1 \approx \sigma_0/2$ and
$\phi_1 = C \, \sigma_0$. This gives $A = \sigma_0\sqrt{1+2C^2\over
2}$. Inflation in this model ends up at
$\phi_e \approx A \sim \sqrt{6\omega} \chi_{e}$ \cite{ExtChaot}.
Consequently, the total increase of the size of the Universe at this
stage of inflation is given by
  \begin{equation}\label{n11}
 {a(t_e)\over a(t_1)} \sim {\left({\phi(t_e) \over
\phi_1}\right)}^\omega \sim {\left(1 +{1\over
2C^2}\right)}^{\omega/2}  \ .
\end{equation}
Thus, for large $\omega$ this stage of inflation can be very long,
and
perturbations generated at this stage will be responsible for the
formation of
the observable part of the Universe. The value of the Planck mass
after inflation is given by
 \begin{equation}\label{6xx}
M_p=  \sigma_0~{\sqrt {\pi(1+2C^2) \over \omega}}~ =
m ~{\sqrt {\pi(1+2C^2) \over \omega\lambda}}~.
\end{equation}

Perturbations of the field $\phi$ at the end of inflation at $\omega
\gg 1$ are orthogonal to the
classical trajectory of the fields $\phi(t), \chi(t)$  in the
$(\phi,\chi)$ space.
Therefore the main contribution to density
perturbations is given by perturbations of the field $\chi$. The
standard
calculation  gives ${\delta\rho\over \rho} \sim {m\over M_p}$, as in
the
ordinary theory $m^2\chi^2$ without any modification of general
relativity.
However, in our case the Planck mass is not a constant, but is
related to $m$
by eq. (\ref{6xx}). This gives ${\delta\rho\over \rho} \sim
{\sqrt{\lambda\omega \over  1+2C^2}}$. In different exponentially
large parts of the
Universe this quantity takes different values corresponding to $1< C
< \sqrt\omega$.

Taking into account our bounds on $C$, we see that from the point of
view of
density perturbations  this model does not exhibit any improvement
as
compared with the usual theory $\lambda\chi^4$, where
${\delta\rho\over
\rho} \sim \sqrt{\lambda}$, and one needs to have $\lambda \sim
10^{-13}$ to satisfy all observational constraints \cite{MyBook}. In
our model the best situation occurs in those domains where quantum
fluctuations lead to $C  \sim \sqrt\omega$, in which case  inflation
is long and ${\delta\rho\over
\rho} \sim \sqrt{\lambda}$. Thus, one still needs to have $\lambda
\leq 10^{-13}$ to satisfy all observational constraints. Before
considering this problem and its
possible
resolution, let us discuss some distinctive features of this model
and some
lessons which we learned when we were developing it.

First of all,  inflation in this theory is possible for all values of
the parameter $\sigma_0$; any potential  $V(\sigma) =  {1\over
4\lambda}(m^2- \lambda\sigma^2)^2$ leads to inflation at small
$\sigma$.
In this respect the  model we consider differs from the analogous
model
in the context of the Einstein theory of gravity, where inflation at
small $\sigma$ is possible only  if  $\sigma_0 >
M_p$. Of course, we do not make any
miracles:  in our model the value of the effective Planck mass
after inflation appears to be  smaller than
$\sigma_0$, see eq. (\ref{6xx}). The subtle but important difference
is that in the Einstein theory
inflation at small $\sigma$ occurs only in the subclass of the models
in which
the two parameters $\sigma_0$ and $M_p$  happen  to be related to
each other
in the above mentioned way, whereas in the context of our model
inflation at
small $\sigma$   occurs for all values of parameters. Moreover, as we
already mentioned, inflation in our model does not suffer from the
problem of initial conditions. This  makes the  existence of the
inflationary regime more robust.

Another interesting feature of this model is the formation of
different domains
of the Universe with different values of the Planck mass and,
correspondingly,
with different amplitudes of density perturbations. According to our
results,  the range of possible variations of $M_p$ and
${\delta\rho\over
\rho}$ is not very wide. However, in our investigation we considered
only the regime when the field $\sigma$ originally was small, $\sigma
\ll \sigma_0$. Whereas at large $\xi$ this was a reasonable
assumption, at small $\xi$ one should consider all other
possibilities as well, including inflation beginning at very large
$\sigma$. Indeed, at small $\xi$ we do not have any unavoidable
symmetry restoration in  our model. Thus, at small $\xi$ our  model
looks like a hybrid of the Brans-Dicke theory and  chaotic inflation
with the potential $V(\sigma)$ which at large $\sigma$ behaves as
${\lambda\over 4}\sigma^4$. In this case there is no upper limit on
possible initial values of $\sigma$  and on the resulting Planck mass
$M_p$, whereas the amplitude of density perturbations in the limit of
large $M_p$ does not depend on $M_p$ and is proportional to
$\sqrt{\lambda}$. The process of self-reproduction of the Universe in
this scenario divides the Universe into many exponentially large
domains where all possible values of $M_p$ are represented
\cite{ExtChaot}.

The situation changes even more dramatically if one considers a
hybrid of  the Brans-Dicke theory and the simplest chaotic
inflation model with $V(\sigma)
= {m^2\sigma^2\over 2}$\, \cite{ExtChaot}. In this case the Universe
becomes divided into exponentially large
domains with the values of the Planck mass taking all values from $m$
to
$\infty$, and with ${\delta\rho\over \rho} \sim {m\over M_p}$ varying
from
O(1) to 0. This opens a very interesting possibility of relating to
each other
the large value of the Planck mass and the small value of
${\delta\rho\over
\rho}$ in our part of the inflationary Universe \cite{BLL}.

In the absence of any realistic model of elementary particle
interactions on
the energy scale discussed in the present paper, it is very hard to
tell
whether the models we are discussing are natural and realistic.
However, it is very encouraging that by making simple hybrids of
basic inflationary models one can obtain an extremely reach variety
of inflationary theories with interesting and sometimes even very
unusual properties. We believe that this enhances the possibility of
finding a correct description of the observational data within the
context of inflationary cosmology.

The author is grateful to J. Bond, E. Copeland, J. Garcia-Bellido, L.
Kofman, D.
Kalligas, A. Laycock, A. Liddle and P. Steinhardt for many
enlightening
discussions, and to D. Linde for the help with computer calculations.
This research was supported in part  by the National
Science Foundation grant PHY-8612280.

\end{document}